\documentclass[12pt,preprint]{aastex}

\shorttitle{Jet precession and X-ray bubbles in NGC~1275}
\shortauthors{Falceta-Gon\c calves et al.}

\begin{document}


\title{Precessing jets and X-ray bubbles 
from NGC~1275 (3C\,84) in the Perseus galaxy cluster: {\it a view from 3D numerical 
simulations}}


\author{D. Falceta-Gon\c calves{1}, A. Caproni{1}, Z. Abraham{2}, D. M. Teixeira{2} and E. M. de Gouveia Dal Pino{2}}
\altaffiltext{1}{N\' ucleo de Astrof\'
isica Te\' orica, Universidade Cruzeiro do Sul - Rua Galv\~ ao Bueno
868, CEP 01506-000, S\~ao Paulo, Brazil}
\altaffiltext{2}{Instituto de Astronomia, Geof\'\i sica e Ci\^encias Atmosf\'ericas, 
Universidade de S\~ao Paulo, Rua do Mat\~ao 1226, CEP 05508-900,
S\~ao Paulo, Brazil}
\email{diego.goncalves@cruzeirodosul.edu.br}


\begin{abstract}

The Perseus galaxy cluster is known to present multiple and misaligned pairs of cavities seen 
in X-rays, as well as twisted kiloparsec-scale jets at radio wavelengths; 
both morphologies suggest that the AGN jet is subject to precession. In this work we 
performed 3D hydrodynamical simulations of the interaction between a 
precessing AGN jet and the warm intracluster medium plasma, which dynamics is coupled to 
a NFW dark matter gravitational potential. The AGN jet inflates cavities that become 
buoyantly unstable and rise up out of the cluster core. We found that under certain 
circumstances precession can originate multiple pairs of bubbles. For the 
physical conditions in the Perseus cluster, multiple pairs of bubbles are obtained for a 
jet precession opening angle $> 40^{\circ}$ acting for at least three precession 
periods, reproducing well both radio and X-ray maps. Based on such conditions, assuming 
that the Bardeen-Peterson effect is dominant, we studied the evolution of the precession 
opening angle of this system. We were able to constrain the ratio between the accretion 
disc and black hole angular momenta as $0.7 - 1.4$. We were also able to constrain the 
present precession angle  to $30^{\circ} - 40^{\circ}$, as well as the approximate age of 
the inflated bubbles to $100 - 150$ Myrs.

\end{abstract}





\section{Introduction}

The powerful radio source 3C\,84 is associated to the elliptical galaxy NGC\,1275 ($z=0.018$), the dominant 
member of the Perseus cluster. Besides its very complex radio morphology at parsec and 
kiloparsec scales characterized by 'S' or 'Z'-shaped jets (Walker et al. 2000, and 
references therein), 3C\,84 
exhibits pairs of kiloparsec X-ray bubbles positioned at different angles with respect to the radio jet. 
The misaligned orientation 
of the bubbles has been interpreted as produced by a jet, precessing with a period of 
$3.3\times 10^7$ years and semi-aperture angle of about 50$^{\circ}$ (Dunn, Fabian \& Sanders 2006, and references 
therein).

The interaction between AGN jets and the intracluster medium (ICM) has already been 
studied in numerical simulations. Scannapieco \& Bruggen (2008) and Bruggen, Sccanapieco 
\& Heinz (2009) studied the evolution of pre-set hot gas bubbles and their role in the 
distribution of energy in galaxy clusters. Heinz et al. (2006) implemented a 
self-consistent procedure by means of AGN jets, though failing in 
inflating bubbles as observed. Typically, the narrow, relativistic and 
nonprecessing jets carve the ICM hot gas and release most of its energy far from the 
central region of the system (Vernaleo \& Reynolds 2006, O'Neill \& Jones 2010), an 
effect known as ``dentist drill". 
Sternberg et al. (2007) showed that wide jets, with opening 
angles $> 50^\circ$, are able to transfer momentum into a larger area resulting in the 
inflation of fat bubbles. The same 
physical properties, however, are expected in precessing AGN jets (Sternberg \& Soker 
2008a). Large precession angles sustained for long periods can be able to inflate 
cavities as observed in the Perseus galaxy cluster.

In this work we present a number of hydrodynamical 3D numerical simulations taking into 
account the evolution of precessing jets, and study the inflation of cavities that can
reproduce the 
observed emission maps of NGC~1275 at radio and X-rays. We assume in this work that the jet 
precession evolution is due to the Bardeen-Peterson effect. The model is described in 
Section 2, as well as the numerical setup. In Section 3 we present the main 
results and the comparison between the synthetic and observed features in NGC~1275, 
followed by the conclusions.

\section{The model}

The basic idea of this work relies on the fact that a precessing jet, originated in the central region of NGC~1275, 
interacts with the intracluster gas resulting in the 
inflation of bubbles. These may rise outward 
of the cluster core, while new bubbles are constantly formed and grow at different 
position angles due to jet precession. 

Several processes may lead to jet precession, such as magnetic torques, warped discs and 
gravitational torques in a binary system (Pizzolato \& Soker 2005).
In this work, we attributed jet precession to the Bardeen-Petterson torques acting on the 
viscous accretion disc that originates the jet, which must be tilted with relation to the 
equatorial plane of the central Kerr black hole (Bardeen \& Petterson 1975). In fact, 
since the origin of the accretion disc is probably related to merging processes, it is 
very unlikely that the BH and the infalling material momentum are perfectly aligned, 
making the Bardeen-Peterson effect a probable precessing mechanism in such an 
environment. 
This effect also induces the alignment of the disc and the black 
hole angular momenta, explaining the presence of twisted jets in several observed AGNs 
(Liu \& Melia 2002, Caproni et al. 2004, 2006, 2007, Fragile \& Anninos 2005, Martin, 
Pringle \& Tout 2007, Chen, Wu \& Yuan 2009).


Under the approximation of a constant surface density accretion disc, Scheuer \& 
Feiler (1996) showed that, for small misalignment angles, its time 
evolution results in an exponential decay of the 
precession angle  $\varphi (t)$ in timescales of the order of 
the precession period. However, King et al. (2005) pointed out that 
this behavior 
is not universal, and more complex time evolutions could be obtained for different disc 
structures. Assuming the total angular momentum vector, defined as 
${\bf J}_\mathrm{T}={\bf J}_\mathrm{BH}+{\bf J}_\mathrm{D}$, being respectively 
${\bf J}_\mathrm{D}$ 
and ${\bf J}_\mathrm{BH}$ the disc and black hole angular momenta, is fully conserved, 
King et al. (2005) derived the equation that rules the time dependency of $\varphi (t)$:

   \begin{eqnarray}
      \frac{d}{d\tau}(\cos\varphi)=\pm \sin^2\varphi\sqrt{\left(\frac{J_\mathrm{T}}{J_\mathrm{BH}}\right)^2-\sin^2\varphi}, 
   \end{eqnarray}
where $\tau=t/T_\mathrm{prec}$, 
$T_\mathrm{prec}$ is the precession period and 
the  '+' and '-' signs correspond, respectively, to the alignment and counter-alignment 
of the involved angular momenta vectors. Notice that $\varphi = \arccos[({\bf 
J}_\mathrm{BH} \cdot {\bf J}_\mathrm{D})/(J_\mathrm{BH} J_\mathrm{D})]$ is the angle 
between the disk and the black hole angular momenta, while $\varphi_\mathrm{D} = 
\arccos[({\bf 
J}_\mathrm{T} \cdot {\bf J}_\mathrm{D})/(J_\mathrm{T} J_\mathrm{D})]$ represents the 
angle 
between the disk and the total angular momenta. Since ${\bf J}_\mathrm{T}$ is constant, 
from Eq.(1) it is possible to obtain the time evolution of both $\varphi$ and 
$\varphi_\mathrm{D}$.


\subsection{Setup of the numerical simulations}

In order to simulate the inflation of the observed cavities detected in X-rays, as well 
as the jet geometry observed at radio wavelengths, we performed
a number of hydrodynamical simulations that provide the evolution of the jet 
interaction with the ICM, as well as the formation of the bubbles.

The model was implemented in a
well-tested Godunov scheme, in which we integrate the full set
of hydrodynamical equations in conservative form (Falceta-Gon\c calves,
Kowal \& Lazarian 2008, Burkhart et al. 2009 and Le\~ao et al. 
2009). The radiative cooling module was computed independently, as we calculate
$\frac{\partial P}{\partial t} = (1-\gamma) n^2 \Lambda(T)$ after
each timestep, where $n$ is the number density, $P$ is the gas pressure, $\gamma$ the 
adiabatic constant and $\Lambda(T)$ is the interpolation function from an electron 
cooling efficiency table for an optically thin gas (Gnat \& Sternberg 2007).

The external gravity was introduced through a
fixed distribution of dark matter following the NFW profile (Navarro, Frenk \& White 
1996),\begin{equation}
\rho_{\rm DM}(r)=\frac{\rho_s}{(r/r_s)(1+r/r_s)^2},
\end{equation}
\noindent where $r_s$ represents the characteristic radius of the
cluster, $\rho_s = M_s/(4\pi r_s^3)$ the mass density, and $M_s$ the absolute mass 
within the radius $r_s$. From isothermal pressure equilibrium the gas density may be 
described by $n(r)=n_0
[\cosh(r/r_s)]^{-1}$. This equation represents the initial density distribution for all 
runs. The temperature was initially set as uniform, being $T_0=10^7$ K. 
As initial setup for the simulations we used an initial core density $n_0=5 \times
10^{-2}$cm$^{-3}$ and $r_s = 30$ kpc, which resulted in a 
profile similar to the empirical density distribution (Sanders et al. 2004).
The computational domain corresponded to three dimensional (3D) cubes with physical size 
$L = 100$ kpc in each direction. The cubes were homogeneously divided into fixed 
256$^3$ and 512$^3$ cells, the later corresponding to $\sim 0.2$ kpc/cell. We used 
open boundary conditions in order to allow gas motion in/outwards the
computational domain, as the pressure gradients evolved with time. 

   \begin{figure}\center
	  {\includegraphics[scale=0.5]{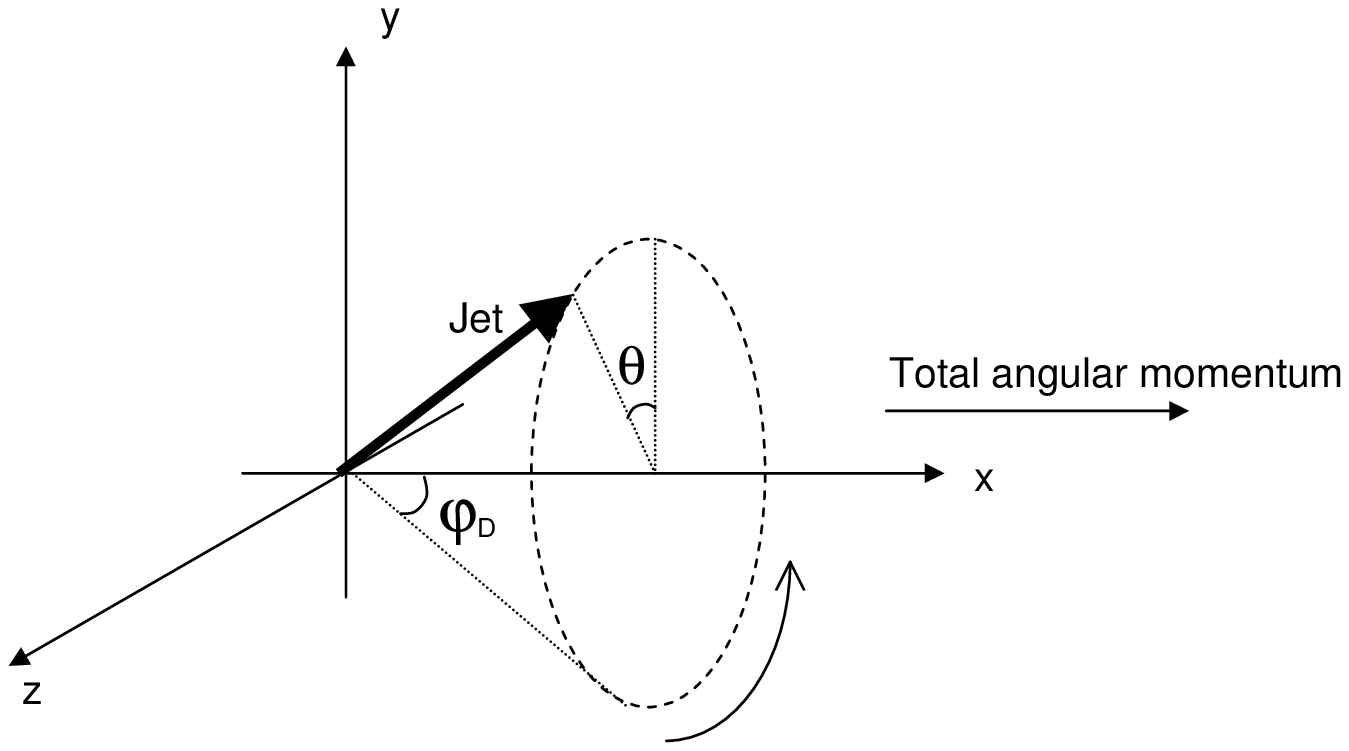}}
      \caption{Scheme used in the simulations for the jet launching. The total angular 
momentum of the system, i.e. the precession axis, is chosen to lay in the x direction. 
The jet precesses around the x-axis with an opening angle $\varphi_\mathrm{D}$ and an 
angular velocity $\dot{\theta} = 2 \pi T_{\rm prec}$. We set $\theta_0 = 0$ as initial 
setup.}      
  \label{scheme} 
  \end{figure}

To simulate the jets we selected a central region in the computational domain with a 
radius of 1kpc with fixed physical conditions.  The local
velocity was fixed as $v_{\rm jet} = 20 c_s$ (which corresponds to $\sim 10^4$ km s$^{-1}$) 
in both opposite directions of the jet that forms an angle $\varphi_D (t)$ with 
respect to the x-axis of the cube, which coincided with the total angular momentum of 
system. The geometry of the simulated jet is cylindrical, i.e. the velocity field of the 
launched jet is uniform. The temperature is set as $T_{\rm jet} = 
10 T_0$, and the density $n_{\rm jet} = 0.1 n_0$, 
resulting in a total mass loss rate of $\sim 0.2$ M$_{\odot}$ yr$^{-1}$ and a total 
kinetic power of $L_{\rm kin} = 10^{43}$ erg s$^{-1}$.
The assumed geometry is shown in Fig.\ 1. The total angular 
momentum of the system $J_\mathrm{T}$, i.e. the precession symmetry axis, is set in the x 
direction. The jet precesses around the x-axis with an opening angle $\varphi_\mathrm{D}$ 
and an angular velocity $\dot{\theta} = 2 \pi T_{\rm prec}^{-1}$. Initially, we set 
$\theta_0 = 0$.

In the simulations, the precession period is assumed 
to be constant as $T_{\rm prec} = 5\times 10^7$yr, close to the value of $3.3 \times 
10^7$yr obtained by Dunn, 
Fabian \& Sanders (2006). We run several models for different initial precession angles 
$\varphi_D(0)$ to study its role on the formation of bubbles. 

As discussed above, the behavior of $\varphi_D (t)$ depends mostly on the 
ratio of the angular momenta of the disc and  BH. Since this value is unknown, we have initially 
assumed in the simulations an alignment law $\varphi_D(t) \propto \exp(-t/\tau^*)$, 
following Scheuer \& Feiler (1996),  and 
varied $\tau^*$, according to the values in Table 1, except for Model \# 13 for which we 
have obtained the precession angle self-consistently from Eq.\ 1.

\begin{table}
\begin{center}
\caption{Description of the simulations}
\begin{tabular}{ccccc}
\hline\hline
Model & resolution & $\varphi_D(0)$ & $^a \tau^{*}$ & Output \\
\hline
1 - 4 & $256^3$ & 10$^{\circ}$ & 0.5, 1.0, 2.0 \& 4.0 & single pair of bubbles\\
5 - 8 & $256^3$ & 30$^{\circ}$ & 0.5, 1.0, 2.0 \& 4.0 & single pair of bubbles\\
9 - 11 & $256^3$ & 60$^{\circ}$ & 0.5, 1.0 \& 2.0 & single pair of bubbles\\
12 & $256^3$ & 60$^{\circ}$ & $4.0$ & multiple pairs of bubbles\\
13 & $512^3$ & 58$^{\circ}$ & $\sim 3.5^b$ & multiple pairs of bubbles\\
\hline\hline
\end{tabular}
\tablenotetext{a}{\ the value of $\tau = t/T_{\rm prec}$ at which 
$\varphi_D(\tau^*)$ = $\varphi_D(0)/e$, assuming $\varphi_D \propto exp(-t/\tau^*)$ - 
except for Model \#13.}\tablenotetext{b}{\ for this model, a self-consistent solution of 
Eq.1 was obtained for $\varphi_D(t)$, giving an approximate 
alignment timescale.}\end{center}\end{table}

   \begin{figure*}[tbh]
      \center
	 \includegraphics[scale=1.2]{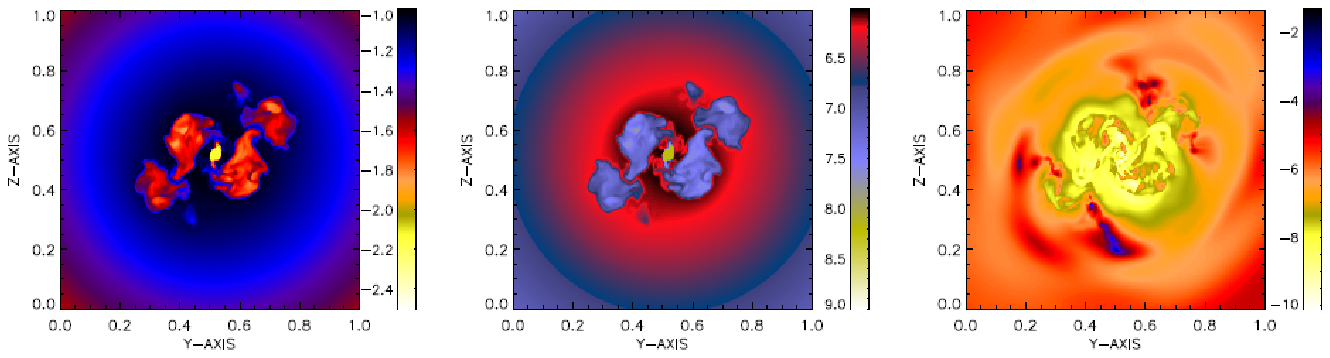}
      \caption{Central Y-Z slices for density (left), in cm$^{-3}$, temperature (center), 
in K, and kinetic energy (right), in erg.cm$^{-3}$, at $t=3 T_{\rm prec}$. 
The plots are shown in logarithmic scales. The total length of the box in each direction 
corresponds to $L = 100$ kpc.}      		
	\label{map1} 
     \end{figure*}

\section{Results}

From the simulated data, obtained under the assumption of exponential decay for 
$\varphi_D$, assuming the gas to be optically thin we constructed  X-ray emission maps 
for different lines of sight. In this case, the X-ray 
emission can be represented by the emission measure, defined as the integral of the 
squared density along the line of sight ($EM=\int n^2 ds$), and directly compared to the 
observations.  Once the synthetic emission maps for all models were calculated, it was 
possible to visually recognize the morphological structures formed up to $t=4T_{\rm 
prec}$. 

As 
main result we found that only models with $\varphi_D(0) \sim 60^{\circ}$ and $\tau^{*} 
> 3.0$ can reproduce the X-ray features in NGC~1275, namely the 
detached and misaligned double pair of bubbles/cavities. Lower values of $\varphi_D(0)$ 
result in single pairs of bubbles, being their widths proportional to $\varphi_D(0)$. 
Also, models with $\varphi_D(0)=60^{\circ}$ but $\tau^{*}<3.0$ produce wide cavities but, 
since the precession is damped in a timescale shorter than the rise time of the bubbles 
only one pair of cavities is seen. With this in place, we were able to obtain the 
ideal set of angular momenta for the disc and the BH, based on Eq.(1). The result has 
been used for the precession angle evolution of Model \# 13.

As the simulation begin, the radiative cooling is responsible for a reduction of 
thermal pressure at the central region where the gas is denser.  A small inward flux 
({\it cooling flow}) is observed (see Falceta-Gon\c calves et al. [2010], 
for details). At later stages, as more energy is released by the AGN, this 
flow is reversed and an outward flux of $\sim 10$km s$^{-1}$ is seen at $t=3T_{\rm 
prec}=150$Myr. The temperature gradient of the ICM, however, is still present, as 
discussed below.

In Fig.\ 2 we show Y-Z central slices of model \# 13 at evolutionary time of $t=3 
T_{\rm prec}$. The density cut (left pannel) clearly shows the existence of two 
disconnected bubbles. The density profile across the bubble shell reveal a density 
contrast $\rho_{\rm peak}/\rho_{\rm ICM} \sim 1.7$, similar to the values obtained by 
(Sternberg \& Soker 2009). This ratio corresponds to $2.6$ at $t=2 T_{\rm prec}$, and 
$3.26$ at $t=1 T_{\rm prec}$. Adiabatic strong shocks result in density 
enhancements of $\sim 4$. This factor can be even larger if cooling is fast. 
At early stages, the jet shock with the ICM gas is supersonic. As the bubble 
expands the shocks become weaker and the density contrast is reduced. 

The temperature map 
(center) is directly correlated 
to the density map. Here, the radiative cooling is responsible for the low 
temperature at the core ($T_{\rm core} \sim 10^6$ K). The cooling timescale $\tau_{\rm 
cool} \simeq 2n_{\rm 0} k_{\rm B} T_{\rm 0} / 3\Lambda(n_{\rm 0},T_{\rm 0}) \sim 75$ Myr 
gives a central temperature $T \sim 10^6$ K after 150 Myr, in agreement with the 
simulations. Within the bubbles, the temperature 
of the low density gas decreases mostly due to adiabatic expansion as $T \sim 2 \times 
10^{42} (t R_{\rm bub})^{\gamma -1}$. The bubbles shown in Fig.\ 2 present a radius 
$R_{\rm bub} \sim 10 - 15$ kpc, which gives $T \sim (2 - 6) \times 10^7$ K. For $r > 
15$ kpc, the cooling has not have enough time to decrease 
the gas temperature. This positive temperature gradient, together with a decreasing 
expansion velocity of the bubble, results in a sharp 
decrease of the Mach number of the shock fronts. A consequence is the appearance of sound 
waves instead of sharp shock fronts. These waves can be recognized as archs in the 
kinetic energy map in Fig.\ 2 (right), and were detected by Graham et al. (2008).

   \begin{figure*}[tbh]
      \center
	 \includegraphics[scale=1]{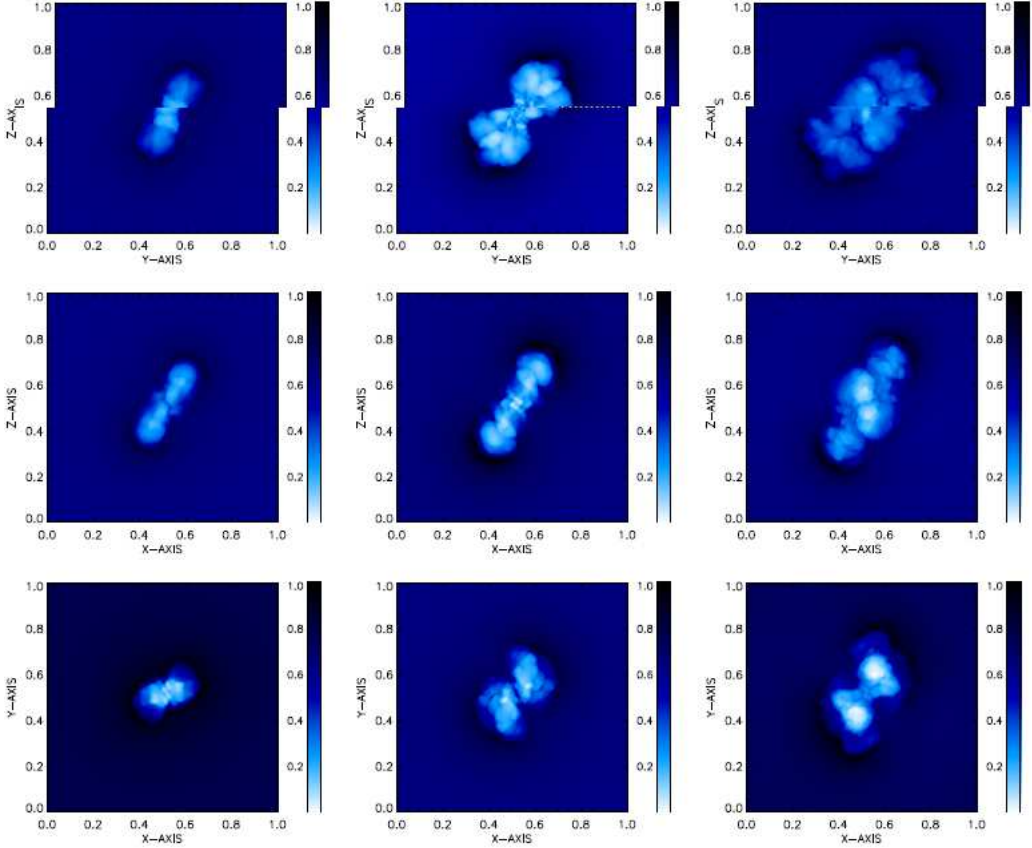}
      \caption{Emission measure ($EM=\int n^2 ds$) maps for the density distribution 
obtained from Model \#13 at evolutionary times $t=1$ (1$^{\rm st}$ column), 
$t=2$ (2$^{\rm nd}$ column) and 
$t=3 T_{\rm prec}$ (3$^{\rm rd}$ column). The integration occurs along x (first 
row), y (center row) and z directions (third row). The total length of the box in each 
direction corresponds to $L = 100$ kpc.}
      \label{map1}   
   \end{figure*}

In Fig.\ 3 we present the emission maps obtained for three different 
lines of sight - along x (first row), y (center row) and z directions 
(third row). Columns 1-3 represent the evolutionary stage at $t= 1$,  $t=2$ 
 and 
$t=3 T_{\rm prec}$.  
The emission measure was normalized 
by its maximum value calculated for each frame. It is noticeable the difference in 
observed morphology depending on the orientation of the line of sight. It occurs because 
of the twisted shape of the cavities due to the precessing jets. The inflated bubbles 
typically detach from the rest of the forming structure at $t \sim (1 - 2) T_{\rm prec}$.

The simulated jets inflated the bubbles in timescales of $\sim 70$ Myrs when they 
detach, at a distance $r_{\rm max} \sim 15$kpc away from the core. At this 
specific time, the cavities present an average diameter $D_{\rm bubble} \sim 25 - 30$ 
kpc. This value is in a good agreement with the analytical approximation for jet inflated 
bubbles (see Eq.\ 3 in Soker [2004]).

Compared to previous numerical simulations of the evolution of rising bubbles, where artificial bubbles are introduced 
in the ICM, (e.g. Bruggen, Sccanapieco \& Heinz 2009), the cavities created 
self-consistently in our simulations are stable for longer periods. Basically, the jet is 
able to continuously input energy into the bubbles, at least 
until detachment. Also, the rising speed of a continuously inflatted bubble is smaller, 
reducing the Rayleigh-Taylor instability effect (Sternberg \& Soker 2008b). 

\subsection{Simulations versus observed data}

As discussed above, using a simple model of exponential decay for $\varphi_D (t)$, it was 
possible to constrain the empirical thresholds of $\varphi_D(0) \sim 60^{\circ}$ and 
$\tau^{*} > 3.0$, in order to obtain two misaligned pairs of bubbles. However, to compare 
with the observed maps, a more detailed evolution of the precession is needed. Using 
Eq.\ 1, we determined the basic physical parameters of the system that lead to the given 
thresholds for the initial precession angle of the jet, as well as the decaying timescale. 
It was possible to reproduce the needed $\varphi_D(0)$ and 
$\tau^{*}$ for $\varphi(0) = 110^{\circ} - 130^{\circ}$ and $J_\mathrm{D}/J_\mathrm{BH}=0.7 - 
1.4$, in Eq.\ 1. For smaller values of $\varphi(0)$ it was not possible to obtain $\varphi_D(0) > 
50^{\circ}$, while for $\varphi(0)>130^{\circ}$  counter alignment occurred instead. For 
each value of $\varphi(0)$ the ratio $J_\mathrm{D}/J_\mathrm{BH}$ determines the 
alignment timescale ($\tau^{*}$), which is well constrained.
From Eq.\ 1 we obtained the variation of $\varphi$ as a function of 
time calculated from Eq.\ 1, assuming $\varphi(0)=120^{\circ}$ and 
$J_\mathrm{D}/J_\mathrm{BH}=1.1$ - the best set of parameters based on the threshold 
given by the simulations. The tilt angles 
$\varphi_\mathrm{D}$ and $\varphi_\mathrm{BH}$ are also obtained, which are respectively 
the angles between $J_\mathrm{D}$ and $J_\mathrm{BH}$ in relation to the x-direction 
defined by $J_\mathrm{T}$. 

A new simulation (Model \#13) was run using a finer resolution 
($512^3$ cells) and the time evolution of $\varphi_D$ given by Eq.\ 1. 
In order to best reproduce the observed X-ray and radio maps we varied the orientation of 
the line of sight and calculated the emission measure, to account for the X-ray synthetic 
map, and integrated the temperature distribution to track the jet 
and its energetic particles, which could be comparable to the observed synchrotron radio 
maps, if the structure of the magnetic fields is not complex. 
The best match between the observed and synthetic maps occurred for a specific line of 
sight inclined $40^{\circ}$ with respect to the total angular 
momentum of the system.


In Fig.\ 4 we included the 
observed 328 MHz VLA radio map of 3C\,84 (a) and the deep {\it CHANDRA} observations of 
the extensive X-ray emission map surrounding NGC1275 (b) (Fabian et al. 2003), which can 
be compared to the temperature integrated map (c) and the emission measure (d), 
respectively, for this specific line of sight. 
   \begin{figure}[tbh]
   \center
	  \includegraphics[width = 70 mm, height = 84 mm]{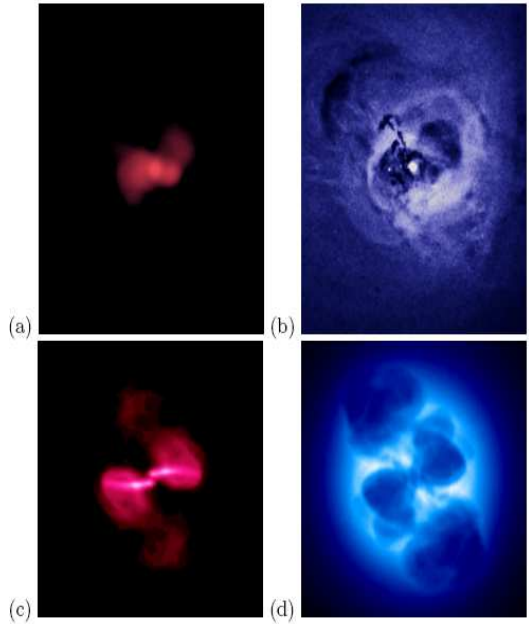} 
         \caption{a) 328 MHz VLA radio map {\it credit NRAO/VLA/G.Taylor}, b) {\it 
credit: NASA/CXC/IoA/A.Fabian}, c) temperature integrated along the line of sight 
normalized by its maximum and d) emission measure normalized by its maximum value. Panels 
$c$ and $d$ correspond to the projection of the mentioned quantities along a line of 
sight inclined $40^{\circ}$ with respect to the total angular momentum of the system. The 
synthetic maps shown were zoomed to better fit the observations. In both cases the total 
length of the image is 70kpc in each direction.}
      \label{maps2} 
  \end{figure}

Visually, the X-ray map is well reproduced by the simulations, except for the fact 
that the farthest pair of bubbles is well aligned in the simulations. In the observed 
maps there is a slight misalignment of the ``ghost cavities". This 
may be related to complex motions of the ICM. The internal cavities 
are in very good agreement with the observations. 
Compared to the radio maps, the projected energy of particles is 
also similar. In this synthetic map the jets seem to be 
misaligned. Actually, this is an effect of 
projection. At the core, the jet points almost towards the observer. The result is 
the formation of two hot spots in the "SW - NE" direction.

This comparison is not straightforward. A self-consistent 
distribution of the magnetic field is required in order to accurately calculate the 
synchrotron radio emission of the jets. Full MHD simulations can be used in the 
near future for this purpose. Qualitatively, we expect the magnetic energy 
to decrease with radius, and the magnetically weighted maps of 
projected particle energy would provide stronger emission closer to the AGN. In 
this case, the map presented in Fig.\ 4c would be more similar to Fig.\ 4a. If 
the simulations truly corresponds to the actual scenario of NGC~1275, 
the timescale used for the synthetic maps, $t=3 T_{\rm prec}$, reveals that 
we may expect a current precession angle $\varphi_D \sim 30^{\circ} - 40^{\circ}$, and an 
evolutionary age of about 100-150 Myrs for the two pairs of bubbles.

\section{Conclusions}

In this work we have studied the role of precessing AGN jets, and the evolution of 
the precession angle with time, on the morphology of the inflated cavities of NGC~1275, 
in the Perseus Cluster. For that, we used a number of hydrodynamical simulations of 
the ICM plasma interacting with precessing jets. We varied the precession angle as well 
as the jet alignment timescales. We found that, in order to reproduce the 
morphology of cavities observed in this system, a precession angle as large as 
$\varphi=60^{\circ}$ is required during a timescale of at least 3 precession periods. In 
such conditions, the jets are responsible for the inflation of two pairs of bubbles, as 
seen in NGC~1275. It is worth to mention the interesting 
agreement between this value and the jet half opening angle predicted by Soker (2004),  
and Sternberg \& Soker (2009) in terms of precessing jets in 2.5D 
simulations. We also found that the condition $v_{\rm jet} \ll c$ is required for the 
inflation of fat bubbles (Sternberg \& Soker [2008a]).

The synthetic emission measure 
obtained from the simulations matches the morphologies of the cavities 
observed in X-rays if the line of sight is inclined by $40^{\circ}$ with respect to 
the total angular momentum of the system. Also, the projected temperature matches 
the synchrotron emission map, though a magnetic field distribution would be required for 
a direct, and more precise, comparison. From the simulations, and assuming the 
Bardeen-Peterson effect to be dominant in this system, we were also able to estimate the 
current precession angle of the AGN jet, $\varphi_D \sim 30^{\circ} - 40^{\circ}$, and 
its age, $t \sim 100 - 150$Myrs.

We were unable to identify from these simulations other complex 
structures observed in 
NGC~1275 such as its filamentary structure detected in H$_{\alpha}$, though this may be 
related to internal turbulence excited by SNe or recent mergers, as shown in 
Falceta-Gon\c calves et al. (2010).

\acknowledgments

The authors thank Prof. Noam Soker for helpful comments that helped improve this 
paper. D.F.G. thanks the financial support of the Brazilian agencies FAPESP (No.\ 
2009/10102-0) and CNPq (470159/2008-1).

\end{document}